# Mechanism Design for Efficient Allocation of Electric Vehicles to Charging Stations

Emmanouil S. Rigas · Enrico H. Gerding · Sebastian Stein · Sarvapali D. Ramchurn · Nick Bassiliades

**Abstract** We study the problem of allocating Electric Vehicles (EVs) to charging stations and scheduling their charging. We develop offline and online solutions that treat EV users as self-interested agents that aim to maximise their profit and minimise the impact on their schedule. We formulate the problem of the optimal EV to charging station allocation as a Mixed Integer Programming (MIP) one and we propose two pricing mechanisms: A fixed-price one, and another that is based on the well known Vickrey-Clark-Groves (VCG) mechanism. Later, we develop online solutions that incrementally call the MIP-based algorithm. We empirically evaluate our mechanisms and we observe that both scale well. Moreover, the VCG mechanism services on average 1.5% more EVs than the fixed-price one. In addition, when the stations get congested, VCG leads to higher prices for the EVs and higher profit for the stations, but lower utility for the EVs. However, we theoretically prove that the VCG mechanism guarantees truthful reporting of the EVs' preferences. In contrast, the fixed-price one is vulnerable to agents' strategic behaviour as non-truthful EVs can charge in place of truthful ones. Finally, we observe that the online algorithms are on average at 98% of the optimal in EV satisfaction.

Emmanouil Rigas · Nick Bassiliades
Aristotle University of Thessaloniki, Thessaloniki , 54124, Greece
{erigas, nbassili}@csd.auth.gr

Enrico H. Gerding · Sebastian Stein · Sarvapali D. Ramchurn
Electronics and Computer Science, University of Southampton,
Southampton, SO17 1BJ, UK
{eg, ss2}@ecs.soton.ac.uk, sdr1@soton.ac.uk

# 1 Introduction

The increasingly negative impact of climate change on society has forced several countries to instigate national plans to reduce carbon emissions [1]. The electrification of transport is one of the main pathways to significantly reduce $CO_2$ emissions. However, the successful introduction of EVs into the market lies upon the acceptance of the new type of vehicle by the customers. Currently, three main problems prevent the spread of EVs: 1) the relatively small range, 2) the long charging times and the unavailability of charging stations, and 3) the higher cost of buying an EV compared to a conventional car. Given that these limitations demand several years before they can potentially be removed, ways of making EVs attractive to customers given the current situation must be developed. For example, the coordinated charging of many EVs given the available stations and the power grid constraints, as well as the fair pricing of the electricity can soften limitations 2 and 3.

In this paper we study a municipality-based urban EV charging setting where EVs drive across a city converge to parking stations in the center and need to charge. The EVs are modelled as self-interested agents that want to charge their battery given a set of preferences and constraints, while from the system's perspective, the maximization of EV satisfaction and the balanced distribution of them across the charging stations and within each station are the objectives. At the same time, the system needs to make a profit and be economically sustainable, but maximizing the profit is not the goal. Hence, it is crucial to leverage advances in decentralized control and mechanism design to coordinate demand and supply to mitigate the impact on the grid.

In order to allocate the EVs to charging points we propose optimal offline solutions, and sub-optimal online ones. In all cases, market-based techniques are being used. In the case of the offline solutions, the EVs report their preferences (e.g., energy demand, time of arrival) a day ahead and the system selects to charge the EVs with the higher valuations given the charging station and network constraints. The valuation is a metric of *how much* the agents want the energy units (i.e., the maximum price that the agent would be willing to pay for an amount of energy). After the allocation has been decided, two pricing mechanisms are used. In the first one, each agent pays a fixed price for each unit of electricity, while in the second one, each agent pays a price calculated by the well-known Vickrey-Clark-Groves (VCG) mechanism [26], [4], [11]. In the case of the online solutions, EV requests are collected by the system over time and the offline algorithms are executed periodically, at pre-defined points over the day, in order to decide the EV-to-charging-station allocation (i.e., a number of requests are being collected and then the scheduling algorithm is called. Thus, each EV waits until the next execution of the scheduling algorithm to learn its allocation).

In this paper, we build on the works by Rigas et al. [19] and Stein et al. [23]. In particular, [19] applies congestion pricing across nodes in a network of charging stations to incentivise EV-agents to charge in stations with low congestion. At the same time, [23] propose a mechanism for allocating electric power units to self-interested agents, aiming to maximise social welfare of the agents. In this paper, we propose a market-based EV to charging station scheduling scheme. This scheme maximizes social welfare from the side of the EVs and guarantees truthful reporting of their preferences, while it minimises charging imbalance both across the charging stations, as well as within each station. Our contributions to the state of the art are:

1. We propose an optimal centralized offline solution using Mixed Integer Programming (MIP) to solve the problem of allocating EVs to charging stations. In doing so, we take into consideration the cost of the electricity, but also an imbalance cost which represents the difference of the actual demand compared to the expected one.
2. We propose two different pricing mechanisms, and we theoretically prove that for one of them truthful reporting of EV preferences is the dominant strategy for all agents.
3. We propose online algorithms that incrementally execute the optimal offline algorithms and achieve near-optimal performance.
4. We experimentally evaluate our algorithms in a setting using real locations of charging stations in Athens, Greece and we show that they achieve high EV satisfaction and have good scalability (i.e., tenths of charging stations and hundreds of EVs).

The rest of the paper is structured as follows: In Section 2 the related work is discussed, in Section 3 the problem is formally defined and in Section 4 the optimal EV to charging station allocation scheme is presented. In Section 5 the two pricing mechanisms are presented, and in Section 6 the online variation of the problem is described. Finally, Section 7 presents the empirical evaluation and Section 8 concludes and presents ideas for future work.

## 2 Related work

Market-based techniques to schedule the charging of EVs across stations have already been applied in a number of occasions [18]. Initial work by Caramanis and Foster [3], investigate market-based control techniques in collectives of EVs for load balancing and to provide regulation services that allow renewable energy sources to be integrated.[1] Specifically, they develop a bidding strategy, using stochastic dynamic programming techniques, for the aggregator to account for uncertain demand from the EVs while maximizing regulation service revenues (by efficiently absorbing unpredictable surges of wind energy into the EV batteries). In [5] they further develop a new bidding strategy (using mathematical programming) for the EV aggregator to operate in hour-ahead (real time) markets.[2]

---

[1] Regulation service corrects for short-term changes in electricity use that might affect the stability of the power system. It helps match generation and load and adjusts generation output to maintain the desired frequency.

[2] Real-Time Market is a spot market in which current prices are calculated at five-minute intervals based on actual grid operating conditions.

In the same vein, González Vayá and Andersson [25] use MIP techniques to bid in a day-ahead market having as an objective to minimise charging costs, while satisfying the EVs' demand for electricity. In [10], the same authors go a step further, and they model the bidding strategy as a two-level problem, where the upper-level is in charge of minimizing the aggregator's charging cost (a set of EVs is represented by an aggregator), while the lower-level represents the market clearing (the price on which electricity is sold), where the bids of other participants are not known in advance. Additionally, Yang et al. [28] propose a centralized charging scheduling framework which also considers the load mismatch risk between the day-ahead and the real-time market.[3] The framework is based on the day-ahead prices and on statistical information of the EVs' driving patterns, while the risk-aware day-ahead scheduling is modeled as a two stage stochastic linear problem which is solved using the L-shaped method [24]. Moreover, Soares et al. [21] present a two-stage stochastic programming approach which uses dynamic pricing for the EVs, but also takes into consideration the uncertainty in energy demand from the EVs and supply from the renewable sources. Finally, Perez-Diaz et al. [17] consider a scenario where a number of independent and self-interested EV aggregators participate in the day-ahead market to purchase energy for their clients' driving needs. In this scenario, independent bidding can drive prices up unnecessarily, resulting in increased electricity costs for all participants. Inter-aggregator cooperation can mitigate this by producing coordinated bids. However, this is challenging due to the self-interested nature of the aggregators, who may try to manipulate the system in order to obtain personal benefit. In order to overcome this issue, the authors employ techniques from mechanism design to develop a coordination mechanism which incentivises self-interested EV aggregators to report their energy requirements truthfully to a third-party coordinator. This coordinator is then able to employ a day-ahead bidding algorithm to optimise the global bids on their behalf.

In the works presented so far, an aggregator collects the preferences of a number of EVs and then it bids in the market trying to satisfy the demand. Aside from such mechanisms, others have also been developed to manage congestion, caused by the simultaneous charging of many EVs, at a local level where each EV is considered separately. A common characteristic in all these mechanisms is that the incentives and allocations are set to ensure the agents have, as their best strategy, to reveal their true preferences for charging times and reserve prices. In particular, we note the work of [23], [6] and [7] that use mechanism design techniques to incentivise self-interested EV agents (that hold their owners' utility function) to book charging slots in order to achieve system-wide objectives (e.g., cost reduction, network stability). Specifically, [23] propose a mechanism for allocating electric power units to self-interested agents, aiming to maximize the social welfare of the agents. In order to generate efficient electricity unit allocation decisions, the authors use a modified version of the Consensus algorithm. Moreover, they use the concept of pre-commitment (the mechanism pledges that it will charge the EV by its departure time, but has the flexibility to choose when and at what rate the charging will take place), they prove that their mechanism incentivises truthful reporting of the preferences of the agents. Instead, in [6], agents state time windows within which they will be available to charge, and bid for units of electricity in a periodic multi-unit auction (one auction per time step). In order to ensure truthfulness, the authors developed a mechanism that occasionally leaves units of electricity unallocated, even if there is demand for them. Moreover, in [7], a two-sided market (between charging points and EVs) is proposed. In particular, the agents report their preferences and their value for the electricity and the charging points report their availability and costs, and then they are allocated the slot that maximizes the difference between their value and the sellers' cost (i.e., in case the agent's value is greater than the cost).

---

[3] An entity (e.g., an EV aggregator) buys electricity in the day ahead market based on predictions on the next day's consumption. Then, in the real time market it can buy (or sell) electricity to cover the actual demand. However, real-time markets are more expensive compared to day-ahead ones, and therefore, the amount of energy bought in the real-time market must be minimised.

In addition, congestion pricing has been used to schedule EVs in such a way so that the load and the congestion to be minimised. For example, [9] develop a decentralized solution where EVs react to a price signal broadcast by the utility a day-ahead. In more detail, two alternative tariffs are explored, one where the same price profile applies system-wide, and another where different prices can be defined at different nodes. By shifting their charging cycles to minimise cost (solving a constrained Optimal Power Flow problem), the EVs also reduce congestion on the distribution network. This solution mainly balances schedules at individual nodes rather than across the network. However, Rigas et al. [19] and Karfopoulos and Hatziargyriou [14] present solutions that balance the charging across a set of stations. In particular, [19] applies congestion pricing across nodes in the network using pricing functions that are demand-dependent. In this way, the EVs (acting as self-interested agents), automatically schedule themselves to minimise congestion and cost across the network but also at individual charging points. Moreover, [14] formulate the problem as a single-objective, non-cooperative, dynamic game and apply a number of price signals across a set of regions of a distribution network. The authors prove that a Nash-equilibrium can be achieved under the assumption that the EV agents are (weakly) coupled.

Ghosh and Aggarwal [8] present an online pricing mechanism where each charging station decides on the prices that will charge to each EV based on the available energy and the time of the day. The authors show that when the mechanism knows the true valuations of the agents it maximizes social welfare (i.e., utilities for the EVs and profit for the stations). However, they do not guarantee truthtelling from the side of the EVs.

In contrast to the works presented so far, in this paper we propose an algorithm that assigns EVs to charging stations, which apart from the EV satisfaction also takes into consideration the balanced charging of EVs across the stations. Moreover, we develop two pricing mechanisms where one of them makes truthtelling the dominant strategy for all EVs. In other words we achieve both load management and social welfare maximization given the set of EV-agents and the available resources. According to the best of our knowledge, this combination is studied for the first time in the EV-related literature [18]. The model of our problem is described in the next section.

## 3 Problem definition

In our model, we define a set of locations $l \in L$, $L = L_p \bigcup L_{\bar{p}}$ that can either be charging stations (i.e., $L_p$) or not (i.e., $L_{\bar{p}}$). Based on this, we define a transport network as a graph $G = (E, L)$ with nodes $L$ and edges $E$, where $e \in E$ represents the roads and $l \in L$ represents the junctions of the road network. We define discrete time points $t \in T$, $T \subseteq \mathbb{N}$, while we denote the set of EVs as $a \in A$.

Now, $\forall l \in L_p$ there are a number of charging slots $s_l \in \mathbb{N}$. Each charging station $l$ has a charging rate, $c_l \in \mathbb{R}_0^+$ (i.e., energy units/time point transferred from a charger to an EV), and all charging stations have a fixed cost $cost_l^{elec} \in \mathbb{N}$ to pay to the electricity provider for each unit of electricity. Thus, we define an allocation matrix $charge_{a,l,t} \in \{0,1\}$ to represent EV $a$ is charging at charging station $l$ at time point $t$. Moreover, each charging station has an expected demand $dem_{l,t} \in \mathbb{N}$ for each time point, which is assumed to be agreed with the electricity provider in advance. In so doing, a monetary penalty $cost_{l,t}^{imbl}$

$$cost_{l,t}^{imbl} = \left| \sum_a charge_{a,l,t} - dem_{l,t} \right| \times cost^{imbl} \qquad (1)$$

is applied to the stations whenever the actual demand varies from the expected one. The term *imbalance* is a measure of how much the actual demand deviates from the expected one. This imbalance cost is calculated as the sum of the absolute value of the difference between the sum of EVs charging at each time point and the expected demand $dem_{a,l}$ over all charging points multiplied by a fixed cost (see Equation 1). In other words, the closer the actual demand is to the expected one, the lower this cost is and therefore the price penalty that each station will have to pay. Note that such imbalance penalties are common practice in the energy industry [12], [16], [2]. Henceforth, index $a$ refers to the EVs, $l$ to the charging stations, and $t$ to the time points.

In this setting, each EV $a$ has a type which is defined by a tuple $\theta_a = \left\langle d_a, b_a^{max}, b_{a,t}, l_a^{start}, t_a^{start}, l_a^{end}, \tau_a^{prk}, t_{a,l}^{arr}, t_{a,l}^{dep}, b_a^{chrg}, v_a^{elec} \right\rangle$. In more detail, each $a$ has a discharge rate $d_a \in \mathbb{R}_0^+$, a maximum battery capacity $b_a^{max} \in \mathbb{R}_0^+$, and a current battery level at time $t$, $b_{a,t} \in \mathbb{R}_0^+$ both measured in kWh. Moreover, each EV $a$ departs from its source location $l_a^{start}$ at time $t_a^{start} \in T$ and wants to travel to destination $l_a^{end}$ where it needs to park for time $\tau_a^{prk} \in \mathbb{N}$. Given a pair of locations $(l_a^{start}, l_a^{end})$ and the transport network $G$, the shortest route, $r_{l_a^{start}, l_a^{end}} \in R_{l_a^{start}, l_a^{end}}$ from the source to the destination is calculated using Dijkstra's algorithm (i.e., $R_{l_a^{start}, l_a^{end}}$ is the set of all possible routes between points $l_a^{start}$ and $l_a^{end}$). Similarly, the shortest routes $r_{l',l} \in R_{l',l}$ from the sources to all charging points $l \in L_p$ are calculated as well. Every route $r_{l',l}$ has a distance $\delta_{l',l} : R_{l',l} \to \mathbb{N}$ measured in kilometers, a time to travel $\tau_{l',l}^{drive} : R_{l',l} \to \mathbb{N}$, and an amount of energy needed, $\epsilon_{l',l}^{need}(\tau_{l',l}^{drive}, d_a)$. Based on slot availability and on the ability of an EV to reach a charging point with its initial battery level, a set of valid charging points $\Gamma_{a,t} \subseteq L_p$ is defined. Now, every EV is available to charge between $t_{a,l}^{arr} = t_a^{start} + \tau_{a,l}^{drive}$ and $t_{a,l}^{dep} = t_{a,l}^{arr} + \tau_a^{prk}$. Note that $\tau_{a,l}^{drive}$ is measured based on the distance to location $l$ divided by an average speed. Each vehicle also needs to charge a specific amount of energy $b_a^{chrg} \leq b_a^{max}$ at charging point $l$ and it also has a personal valuation $v_{a,l}$ for charging the desired amount of energy at each charging point.

$$v_{a,l} = \begin{cases} v_a^{elec} - \kappa_{a,l}^{time}, & \text{if } b_{a,t_{a,l}^{dep}} \geq b_{a,l}^{chrg} \\ 0, & \text{otherwise} \end{cases} \quad (2)$$

According to Equation 2, a time cost $\kappa_{a,l}$ related to driving to the station and walking from the station to the final destination is subtracted from the valuation $v_a^{elec}$ for charging the desired amount of electricity. Time cost $\kappa_{a,l}$ actually shows how willing agent $a$ is, to drive to point $l$ and then walk to the final destination. Note that, the agent has zero valuation for charging less than $b_a^{chrg}$, and valuation $v_{a,l}$ for charging equal to or more than $b_a^{chrg}$. Moreover, each agent $a$ receives utility $u_a$,

$$u_a = v_{a,l} - p_a \quad (3)$$

where $p_a \in \mathbb{R}$ is a monetary transfer from the EV to the system (i.e., the utility is a measure of *satisfaction* for charging the desired amount of electricity). $p_a$ is usually positive, as the EVs pay the charging station for the electricity. However, in case the charging of an EV leads to lower imbalance cost for the charging station, the transfer for this EV may be negative (i.e., the EV gets paid- see Section 5.2).

For an EV to be assigned to a charging station and charge, it has to reveal its type $\theta_a$ to the system. Then, the system applies an EV to charging station allocation algorithm to schedule EV charging and uses one of two proposed pricing mechanisms to calculate the prices for the EVs. Such mechanisms can be either offline (Section 4), or online (Section 6).

## 4 Offline scheduling of EVs to charging points

Here, we present a formulation for optimally allocating EVs to charging points. Thus, we formulate the problem as a Mixed Integer Programming (MIP) one and we solve it optimally using IBM ILOG CPLEX 12.6.2.

In our formulation we define two decision variables: 1) $\phi_{a,l} \in \{0, 1\}$ which indicates whether an agent $a$ is serviced at charging station $l$ and 2) $charge_{a,l,t} \in \{0, 1\}$ which indicates whether agent $a$ is charging at charging station $l$, at time point $t$ (at charging rate $c_a$). The objective of this formulation is to maximize

Equation 4, which means to maximize the sum of the agents' valuations (i.e., charge as many agents as possible at the stations with the lower time cost) minus the cost for the electricity and minus the imbalance cost for an optimal allocation $X^*$ of EVs to charging points. Note that the absolute value in the objective function is linearized at run time by CPLEX.[4] More formally, the linear program is given by:

Objective function:

$$F(X^*) = \sum_a \sum_l v_{a,l} \times \phi_{a,l} - \sum_a \sum_l \sum_t charge_{a,l,t} \times cost_l^{electr} - \sum_l \sum_t cost_{l,t}^{imbl} \quad (4)$$

Subject to:

$$\sum_l \phi_{a,l} \leq 1, \forall a \quad (5)$$

$$b_a^{arr} \geq \epsilon_{a,l}^{need}(r_{a,l}, d_a) \times \phi_{a,l}, \forall a, \forall l \quad (6)$$

$$\sum_{t \geq t_{a,l}^{arr} \cap t < t_{a,l}^{dep}} charge_{a,l,t} \geq \left[b_{a,l}^{chrg}/c_l\right] \times \phi_{a,l}, \forall a, \forall l \quad (7)$$

$$\sum_{t < t_{i,j}^{arr} \cap t \geq t_{i,j}^{dep}} charge_{i,j,t} = 0, \forall i, \forall j \quad (8)$$

$$\sum_{t \geq t_{a,l}^{arr} \cap t < t_{a,l}^{dep}} charge_{a,l,t} + b_a^{arr} \leq b_a^{max} \quad (9)$$

$$\sum_a charge_{a,l,t} \leq s_l, \forall l \forall t \quad (10)$$

We detail the key elements of this mathematical program as follows. EV $a$ will charge at most at one charging station $l$ (Equation 5), and the initial battery level of each EV must be enough to reach the charging station selected to charge (Equation 6). In addition, the number of time points that an EV will be charging must be greater or equal to the energy demand divided by the charging rate $c_l$ at station $l$ ($\tau_{a,l}^{chrg} = \left[b_{a,l}^{chrg}/c_l\right]$) (Equation 7). Moreover, for all time points before the arrival and after the departure of EV $a$ at charging station $l$, no charging must take place (Equation 8), and after the completion of the charging, the maximum capacity of the EV's battery must not be exceeded (Equation 9). Finally, the maximum capacity of each charging station, in terms of available chargers, must not be violated at any time (Equation 10).

In the next Section, we present two pricing mechanisms for the electricity that each EV charges.

---

[4] This is usually achieved by adding two extra decision variables and two extra constraints.

## 5 Pricing mechanisms

In this section, we present two mechanisms for calculating the price that the agents will pay for the electricity they will charge according to the scheduling as presented in the previous section. In the first mechanism, the agents are assumed to be truthful, while in the second one they may misreport their types.

In order to evaluate a mechanism, we examine whether a number of properties hold:

1. **Individual Rationality**: This means that all agents that are scheduled to charge have non-negative utility, while the ones that are not scheduled to charge have zero utility (i.e., $\forall a, u_a \geq 0$). In our setting, this specifically means that agents will never pay for the energy more than their valuation.
2. **Dominant Strategy Incentive Compatibility**: This means that for all agents, truthfully reporting their types is the best strategy, no matter what the other agents do (i.e., $u_a(\theta_a) \geq u_a(\hat{\theta}_a)$). In our setting, this means that the agents will not end up paying higher prices if they are truthful regarding their valuations for the energy.
3. **Efficiency**: A mechanism, such as the one presented in this work is *efficient*, if in equilibrium it selects a choice $X^*$ such that it maximizes social welfare: $\forall X^*, F(X^*) \geq F(X'^*)$. In our case, this means that the available energy units will be charged to the agents with the higher valuations.
4. **Budget Balance**: This means that the sum of all transfers (i.e., payments to and from the charging stations) are equal to zero (i.e., $budget = \sum_a (p_a^{vcg} - cost_a^{elec}) - \sum_l \sum_t (cost_{l,t}^{imbl}) = 0$). Weak budget balance means that *budget* is non-negative, while no-budget balance means that *budget* can get any value. In our setting, this is needed for the sustainability of the community stations, meaning that the stations should not suffer from losses but also that maximizing profit is not their goal.

### 5.1 Cooperative Agents

Initially, we assume that all agents are cooperative (i.e., they report their types $\theta_a$ truthfully) and we design a fixed price mechanism where the payments of the agents to the charging company are calculated based on:

$$p_a^{coop} = (b_a^{chrg} \times cost_l^{elec}) \times (1 + incr) \qquad (11)$$

Based on this equation, each agent pays its energy demand multiplied by the cost of electricity for each unit of energy increased by a percentage $incr \in \mathbb{R}^+$ (i.e., this value determines the profit that the charging station will make for each unit of electricity that sells to each EV. This is actually the usual price setting mechanism in many markets, where the seller prices a product based on its cost increased by a fixed percentage). In order to calculate the *incr* we find the point where the mechanism becomes sustainable, i.e., stops making losses and starts making a small profit (see detailed description in Section 7). From now on, we will refer to this mechanism as *Coop*.

The allocation of the agents to charging points takes place based on the objective function (Equation 4). However, the price to pay is calculated afterwords and the valuation of the agents is not taken into consideration. Thus, the price to pay can be higher than an agent's valuation. In this case, the agent decides not to charge and receives zero utility. For this reason, we conclude that the agents' utilities are always equal or larger than zero and the mechanism is individually rational.

Under the assumption that all agents are truthful, they will report their true valuations and for this reason the mechanism would be incentive compatible. However, as long as an agent will not get negative utility it has an incentive to misreport its valuation. For example, assuming that the imbalance cost is equal to zero, and $incr = 0.05$, if an agent has valuation $v_{a,l} = 5$ but reports $v'_{a,l} = 6$, and $b_i * cost_l^{electr} = 4$, the optimizer will schedule this agent to charge and the price to pay will be $p_a^{coop} = 4 + 4 * 0.05 = 4.2$. Thus, the agent will charge with the same price, but it will increase the chances of being selected instead of another agent that would report its valuation truthfully. For this reason, the mechanism is not incentive compatible. This will be experimentally confirmed in Section 7.4.

Now, in order the mechanism to be efficient, Equation 4 must be maximized. Under the assumption that all agents are truthful, the optimization procedure leads to an allocation of EVs to charging points which maximizes this function. However, in case for some agents the price to pay is higher than their valuation, these agents will decide not to charge. Thus, some resources will remain unallocated. Also, the assumption that the agents will be truthful does not always hold. For these reasons, the mechanism is not efficient.

Finally, in the case where the actual demand is equal to the expected one, then the mechanism will make a profit as $totalCost = b[i] * cost^{elec} + 0$ and $totalRevenue = b_i * (cost^{elec} + cost^{elec} \times incr) > totalCost$. If the demand is different than the expected one, then the budget can be either positive or negative. Only in the case where $\sum_l \sum_t (cost_{l,t}^{imbl}) = \sum_i (b_i \times cost^{elec} \times incr)$ the mechanism is budget balanced. Thus, in the general case, the mechanism is not budget balanced.

In the next section we present an alternative pricing mechanism which makes truthful reporting of preferences the dominant strategy for all agents.

5.2 Strategic Agents

In the general case, agents would try to misreport their types if they had an incentive to do so (i.e., achieve higher utility). The mechanism presented in the previous section, can easily be manipulated if some agents misreport their type (e.g., report higher valuation). In this section, we present an optimal EV to charging station allocation scheme which uses the well known Vickrey-Clarke-Groves (VCG) mechanism [26] [4], [11]. The VCG mechanism is a generalization of the Vickrey auction where, in the general case, multiple agents bid for multiple goods of the same type (i.e., combinatorial auction) and the price to pay for each agent is calculated based on the harm they cause to the other agents [27]. The main characteristic of this mechanism is that it is incentive compatible, which means that no agent can benefit by declaring anything other than its true type. Therefore, this mechanism, assuming that all agents play their dominant strategies, finds the optimal allocation of the resources (i.e., electricity units) in terms of social welfare maximization and then calculates the price that each agent will pay to the mechanism.

In order to calculate the allocation of EVs to charging points, we use the MIP formulation as described in Section 4. In the end of the optimization procedure, an optimal allocation $X^*$ of EVs to charging points is achieved. We calculate the transfer $p_a^{vcg}$ (i.e., the price) that EV $a$ will pay to the mechanism for the energy charged, as follows:

$$\begin{aligned}
p_a^{vcg} &= (\sum_{e \in A}(v_{e,l}(X^*_{-a}) - cost_e^{elec}(X^*_{-a}) - cost_e^{imbl}(X^*_{-a})))- \\
&\quad (\sum_{e \in A}(v_{e,l}(X^*) - cost_e^{elec}(X^*) - cost_e^{imbl}(X^*)) - v_{a,l}(X^*)) = \\
&\quad \sum_{e \in A}(v_{e,l}(X^*_{-a}) - cost_e^{elec}(X^*_{-a}) - cost_e^{imbl}(X^*_{-a}))- \\
&\quad \sum_{e \in A}(v_{e,l}(X^*) - cost_e^{elec}(X^*) - cost_e^{imbl}(X^*)) + v_{a,l}(X^*) \quad (12)
\end{aligned}$$

Based on this equation, each agent $a$ will pay its impact on the others (i.e., its *social cost*) added to the cost of electricity it charged and the imbalance cost (i.e., $X^*_{-a}$ denotes the optimal allocation without the existence of agent $a$, $v_e(X^*)$ the valuation of agent $e$ based on an optimal allocation $X^*$, $cost_e^{elec}(X^*)$ the electricity cost for agent $e$ based on an allocation $X^*$ and $cost_e^{imbl}(X^*)$ the imbalance cost for $e$ based on $X^*$). In more detail, the first sum contains the total values and costs for all agents, but in an allocation where $a$ does not exist. Whereas, the second sum contains the total value of all agents apart from $a$ and the costs for all agents including $a$ in an allocation where all agents participate. In all cases, the cost of electricity is fixed (per unit of energy) for all agents and acts as a reserve value for the charging station, while the imbalance cost depends on the demand profile. From now on, we will refer to this algorithm as *VCG*.

In contrast to the previous mechanism where agents have an incentive to lie, here due to the fact that 1) VCG mechanism is used and 2) the types of the EVs are not interdependent, it is best for the agents to reveal their types truthfully. In the rest of this section we prove the properties of this mechanism.

**Theorem 1** *The VCG mechanism for the EV allocation problem is individually rational.*

*Proof* The result of the optimization procedure is a set of agents selected to charge $A' \subseteq A : \forall a \in A', v_{a,l} - cost_{a,l}^{elec} - cost_l^{imbl} \geq 0$ and an allocation $X^*$ to charging stations. Now, $\forall a \in A'$, the transfer $p_a^{vcg}$ from agent $a$ to the system is given by Equation 12, and the utility of agent $a$ is given by Equation 3 which, based on Equation 12 becomes:

$$u_a = v_{a,l} - p_a^{vcg} = v_{a,l}(X^*) -$$
$$\sum_{e \in A'}(v_{e,l}(X^*_{-a}) - cost_e^{elec}(X^*_{-a}) - cost_e^{imbl}(X^*_{-a})) +$$
$$\sum_{e \in A'}(v_{e,l}(X^*) - cost_e^{elec}(X^*) - cost_e^{imbl}(X^*)) - v_{a,l}(X^*) =$$
$$\sum_{e \in A'}(v_e(X^*) - cost_e^{elec}(X^*) - cost_e^{imbl}(X^*)) -$$
$$\sum_{e \in A'}(v_e(X^*_{-a}) - cost_e^{elec}(X^*_{-a}) - cost_e^{imbl}(X^*_{-a})) \quad (13)$$

Now, in order the $\sum_{e \in A'}(v_e(X^*) - cost_e(X^*)^{elec} - cost_e^{imbl}(X^*))$ to be greater or equal to $\sum_{e \in A'}(v_e(X^*_{-a}) - cost_e(X^*_{-a})^{elec} - cost_e^{imbl}(X^*_{-a}))$ and therefore, Equation 14 to be true, the *choice-set monotonicity* and the *no negative extrnalities* properties must hold [20]. Choice-set monotonicity means that by removing any agent, the mechanism's set of possible choices weakly decreases. In our case, this property holds as, if an EV leaves, the mechanism has fewer choices in scheduling the EV charging. Moreover, the no negative externalities means that every agent has zero or positive utility for any choice that can be made without its participation. In our case, this property also hods, as if an agent does not participate in the mechanism, it gets utility equal to zero. Also, note that if the second sum was larger than the first one, the optimizer would not select $i$ to charge in first place. For these reasons, Equation 14 always holds and the mechanism is individually rational (Equations 15, 16).

$$\sum_{e \in A'}(v_e(X^*) - cost_e^{elec}(X^*) - cost_e^{imbl}(X^*)) -$$
$$\sum_{e \in A'}(v_e(X^*_{-a}) - cost_e^{elec}(X^*_{-a}) - cost_e^{imbl}(X^*_{-a}))$$
$$\geq 0, \forall a \in A' \quad (14)$$
$$u_a = 0, \forall a \notin A' \quad (15)$$
$$u_a \geq 0, \forall a \in A' \quad (16)$$

We can safely assume that the agent would not lie about the discharging rate, the maximum battery capacity of the EV, the initial battery level, the start location and the final destination. The discharging rate and the maximum battery capacity are considered to be common knowledge. If an agent lies about its initial battery level, then it reduces the options that the scheduler has to assign it to a charging station. At the same time, there is no point in lying about its start and end location as in that case it would have to move to or from these locations, which could cost the agent more than it would be gaining. Given these, we can prove incentive compatibility of the mechanism:

**Theorem 2** *The VCG mechanism for the EV allocation problem is dominant strategy incentive compatible under the assumption that the system knows the discharging rate, the maximum battery capacity, the initial location and the final destination of each EV.*

*Proof* Agents could have an incentive to misreport their energy demand ($b_{a,l}^{chrg}$) and their valuation for this energy ($v_{a,l}$), the time cost ($\kappa_{a,l}$), their arrival and departure times ($t_{a,l}^{arr}$, $t_{a,l}^{dep}$). We assume that the discharging rate, the maximum battery capacity of the EV, the initial battery level, the start location and the final destination are common knowledge to the scheduling centre.

In what follows we will prove incentive compatibility for the agents' reported type, for an optimal allocation $X^*$. The utility of agent $a$, when it reports its type $\theta_a$ truthfully is

$$u_a = v_a - (F(X^*_{-a}) - F(X^*) + v_a) =$$
$$v_a - F(X^*_{-a}) + F(X^*) - v_a =$$
$$= v_a - F(X^*_{-a}) + F_{-v_a}(X^*) \quad (17)$$

while, when agent $a$ reports its type $\hat{\theta}_a$ non-truthfully its utility is

$$\hat{u}_a = v_a - (F(X^*_{-a}) - F(\hat{X}^*) + \hat{v}_a) =$$
$$v_a - F(X^*_{-a}) + F(\hat{X}^*) - \hat{v}_a =$$
$$= v_a - F(X^*_{-a}) + F_{-\hat{v}_a}(\hat{X}^*) \quad (18)$$

In this case, the utility is the difference between the true valuation of agent $a$ and the impact on the other agents based on the non-truthful report of its type.

The first two terms of Equations 17 and 18 are not affected by the non-truthful report of the type of agent $a$ (Reminder, F(X*) is the value of the objective function- Equation 4). Thus, in order to understand whether the mechanism is incentive compatible, we have to see how the final term can be affected. In so doing, we evaluate a number of cases:

1. Agent $a$ is non-truthful and charges:
   (a) If its report does not affect anyone else, then:

   $$F_{-\hat{v}_a}(\hat{X}^*) = F_{-v_a}(X^*) - cost_a^{elec} - cost_a^{imbl}$$
   $$< F_{-v_a}(X^*) \quad (19)$$

   Thus, $\hat{u}_a < u_a$ and for this reason agent $a$ has no incentive to lie.

(b) If its report affects agent $b \in A : b \neq a$, then:

$$F_{-\hat{v}_a}(\hat{X}^*) = F_{-v_a}(X^*) - (v_b - cost_b^{elec} - cost_b^{imbl}) \\ - cost_a^{elec} - cost_a^{imbl} \quad (20)$$

Given that $v_b - cost_b^{elec} - cost_b^{imbl}$ is greater than zero (otherwise $b$ would not have been selected to charge at first place), $F_{-\hat{v}_a}(\hat{X}^*) < F_{-v_a}(X^*)$ and for this reason $\hat{u}_a < u_a$. Therefore, $a$ has no incentive to lie. Note that if this agent's report affected more than one agent, its loss would be even greater.

2. Agent $a$ is non-truthful and does not charge. In this case, $\hat{u}_a = u_a = 0$. Thus it doesn't have an incentive to lie.

Note that if agent $a$ reports $\hat{t}_a^{arr} < t_a^{arr}$ or $\hat{t}_a^{dep} > t_a^{dep}$ then the optimizer could schedule it to charge at time points that it wouldn't be at the charging station. Thus, the agent would get less energy compared to its demand and, therefore zero utility. Also, if agent $a$ reports $\hat{b}_a < b_a$, then by default it would get zero utility. Moreover, if $\hat{b}_a > b_a$, the optimization already takes that into consideration (i.e., Constraint 7). So if an agent would be better off reporting higher energy demand (as this could decrease the imbalance cost) then the optimization will do it automatically.

**Theorem 3** *The VCG mechanism for the EV allocation problem is efficient (i.e., maximizes social welfare).*

*Proof* In order the mechanism to be efficient, Equation 4 must be maximized. Indeed, after the optimization procedure, and given that the mechanism is incentive compatible, the allocation of EVs to charging points leads to the maximization of this function. Therefore, the allocation is efficient.

**Theorem 4** *The VCG mechanism for the EV allocation problem is not budget balanced.*

*Proof* The budget of the mechanism is given by:

$$budget = \sum_a (p_a^{vcg} - cost_a^{elec}) - \sum_l \sum_t (cost_{l,t}^{imbl}) \quad (21)$$

In the general case where the actual demand is different from the expected demand, budget balance cannot be guaranteed. In the case of very low demand, the stations will have a loss as the income will be low but the imbalance cost very high. For example, if the expected demand is 2 EVs at each time point, and no EV arrives to charge, the station will have a loss as it will need to pay the imbalance cost as calculated from Equation 1. Moreover, given that the charging of an EV can reduce the imbalance cost, negative transfers are also possible. In the case where the actual demand is higher than the expected one, the EVs will have positive transfers to the stations. Thus, our mechanism is not budget balanced. However, in our setting, although profit maximization is not the main objective, this should not be considered a problem, as the aim of the charging stations should be to make some profit so as to be economically sustainable.

Now, in the extreme case where the actual demand matches exactly the expected demand, then $cost_a^{imbl} = 0, \forall a$. In this case, as the charging of any EV does not affect the charging of another (we assume that the pre-agreed consumption can always be covered by the station). Thus, the sum of all transfers from the agents to the stations is equal to the cost of electricity paid by the stations to the electricity provider ($\sum_a(p_a^{vcg}) = \sum_a(cost_a^{elec}) \Rightarrow budget = 0$). Only in this case, our mechanism is budget-balanced.

## 6 Online scheduling of EVs to charging points

So far, we assume that the demand becomes known to the system a day ahead. However, in this section, we present an online version of the EV to charging station scheduling problem where agents arrive in the system dynamically over time and need to charge. In so doing, the system collects the requests from the agents and clears the market at pre-defined points in time similar to [6]. By market clearing we actually mean that the EVs that have reported their preferences to the system are considered in the EV to charging station scheduling (see also Figure 1 and Algorithm 1). In more detail, a sequence of points $t_p \in T$ in the day where market clearing takes place are defined. After each $t_p$ the charging scheduling algorithm (Section 4), is executed and an optimal allocation $X^*$ is calculated. Later, the price to pay for each EV is calculated based on either the Coop or the VCG mechanism. Now, for every $p : p > 1$, the mechanism makes sure that the already existing schedule of EVs charging is not affected. This is an important assumption, as it will later guarantee incentive compatibility from the side of the EVs in the case where the VCG is used to calculate the prices. Also, note that all EVs that participate at the market at $t_p$, must have an arrival time $t_a^{arr} < t_p$.

---
**Algorithm 1** Online EV to charging station scheduling.
---
**Require:** $A$, $T$, $L$, $\theta_a \forall a$, $dem_{l,t}$
1: **for all** $p$ **do**
2:     call offline algorithm for all $a \in A : t_a^{start} \leq t_p$
3:     {For the EVs that were scheduled in earlier market clearing, the value of $\phi_{a,l}$ remains unchanged. The payments are calculated in the same way as in the offline algorithms.}
4: **end for**
5: **return** $\phi_{a,l}$
---

The properties of the VCG mechanism, as these have been described in the previous section, hold in the online setting as well. However, special notice should be given to the *dominant strategy incentive compatibility* and the *efficiency*:

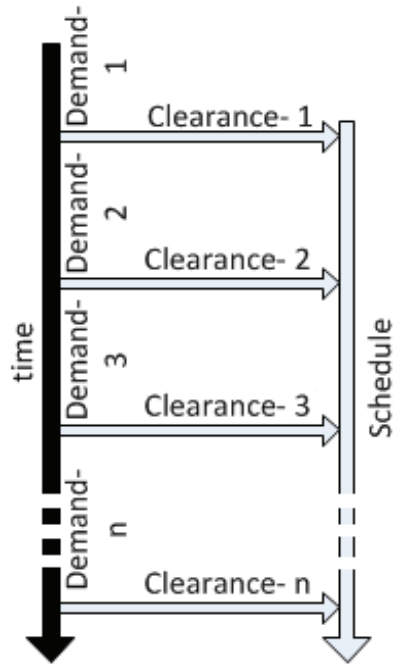

**Figure 1** Online EV to charging station scheduling

Regarding the incentive compatibility, agents could try to misreport their arrival time in order to participate in a later market clearing. However, as agents can report their demand for any time in the day, by delaying the report they cannot have higher utility. This is due to the following reasons: 1) Given the imbalance cost, the earlier an agent arrives to the system, the higher the improvement it causes to the imbalance cost. Thus, the transfer of the agent to the mechanism could become smaller. 2) If an agent delays the report, stations can already be highly congested, thus not being able for it to charge. Therefore, the best strategy for the agents is to report their preferences to the system the earlier possible time (i.e., the time they decide that they want to charge).

Moreover, regarding efficiency, the mechanism is efficient for each market clearing. However, it may not maximize social welfare for the whole set of agents. This is inevitable due to the fact that the system does not know the future demand. Thus, agents with possibly low valuations, that have participated in an early market clearing may have occupied charging slots that could have later been used by agents with higher valuations.

In the next section, we evaluate our algorithms in a realistic setting and for a number of problem dimensions.

## 7 Evaluation

In this section we evaluate our algorithms using real data regarding locations of charging stations and numbers of available plugs in Athens, Greece.[5] The EV demand across the stations (and the valuations) is based on observations of traffic flow from Google maps, and discussions in related works [13], [22]. In more detail, we evaluate the execution time and scalability (EXP1), the EV satisfaction (EXP2), the cost for the EVs and profit for the charging stations (EXP3) and the impact of preferences' untruthful reporting (EXP4). In so doing, we use 50 time points, where each one is assumed to be equal to 15 minutes, 10 - 130 EVs and 8 charging stations. The arrival and departure times of the EVs, the energy demand, the valuation for each energy unit and the expected demand are drawn from uniform distributions (i.e., $t_a^{arr} : mean = 15, \sigma = 15$, $t_a^{dep} : mean = T - t_a^{arr}$, $b_a^{chrg} : mean = t_a^{dep} - (t_a^{dep} - t_a^{arr})/2, \sigma = t_a^{dep} - t_a^{arr})/2$, $v'_{a,l} : mean = 0.5, \sigma = 0.5$ which is then multiplied by the number of energy units the agent wants to charge $v_{a,l} = v'_{a,l} \times b_a^{chrg}$ and $dem_{l,t} : mean = 2, = 1$). Note that the charging rate is fixed to one unit of energy per time point in all stations and that the desired energy is always able to be charged in the EV within the available time window. Finally, the *incr* value for the Coop mechanism is calculated as follows: An initial value of 0.1% is given to *incr* and the optimization is executed recursively each time increasing *incr* by 0.1%. At each iteration some EVs may leave the system as the price to pay gets higher than their valuation. The value of *incr* is fixed to the one where the mechanism starts making a profit. This procedure was executed multiple times for different numbers of EVs and the average value of $incr = 2.5\%$ was selected. Finally, the time points at which the scheduling takes place in the online scenario are $t_p = \{10, 20, 30, 40, 50\}$. Note that for each $p$, requests collected at any $t : t \geq t_{p-1}$ and $t < t_p$ are considered.

### 7.1 EXP1: Execution time and scalability

Highly combinatorial problems such as the one we solve here are known to suffer from high execution times. Thus, it is crucial to evaluate the execution time of both the offline as well as the online algorithms and for a number of scenarios (see Figure 2).

In terms of the VCG mechanism, the execution times for both the offline and the online versions increase quadratically ($R^2 = 0.898$ for the offline and $R^2 = 0.998$ for the online). For up to 60 EVs, both formulations have execution times which are well under 30 seconds. However, later the execution time for the offline version increases rapidly and for 130 EVs it reaches 778 seconds, while the online algorithm executes in 31 seconds. Note that for the online version, we present the average execution time for all market clearings for each number of EVs (reminder: in the online version the scheduling algorithm is called at specific points in time). The online algorithm must calculate a charging schedule fast, and for this reason the low execution time is crucial for its usability. For the 130 EVs, the 45 seconds can be considered as an acceptable value. However, it should be noted that due to the fact that for each market clearing the EVs that have already been scheduled to charge are also considered (as a constraint this time), the optimization procedure remains quite complex. Overall, the second part of the objective function which contains the absolute value, although it is linearized at run time, it affects the execution time a lot.

---

[5] https://user.fortizo.gr/#/portal/locations.

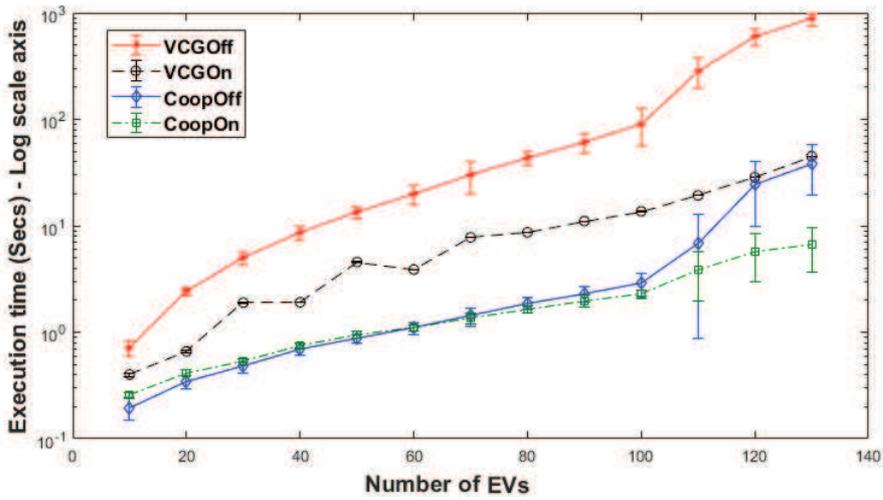

**Figure 2** Execution times of all algorithms

As long as the Coop mechanism is concerned, the execution time grows quadratically for the offline ($R^2 = 0.85$) and the online version ($R^2 = 0.956$). However, the rate of growth is considerably smaller compared to the equivalent times of the VCG mechanism. For example, for 130 EVs, the execution time for the offline version is approximately 38 seconds and for the online one 6.69 seconds. Note that the error bars in all graphs show the standard deviation of each sample.

7.2 EXP2: EV satisfaction

In terms of the average number of serviced EVs (Figure 3), for up to 100 EVs on average 85% of all vehicles are charged. However, for more than 80 EVs more of them remain uncharged, as for example for 130 EVs 81.3% of them are charged by the VCG offline and 79.6% by the Coop offline. Actually, the offline mechanisms always lead to slightly higher numbers of serviced EVs. This was expected due to the fact that when the problem is solved online, congestion and resource management are not optimal and for this reason some EVs remain uncharged. When comparing the two mechanisms, VCG always services slightly more cars than the Coop. This can be explained by the fact that the Coop does not take into consideration the valuations of the EVs in setting the prices. Thus, some EVs get higher price than their valuation and they leave without charging.

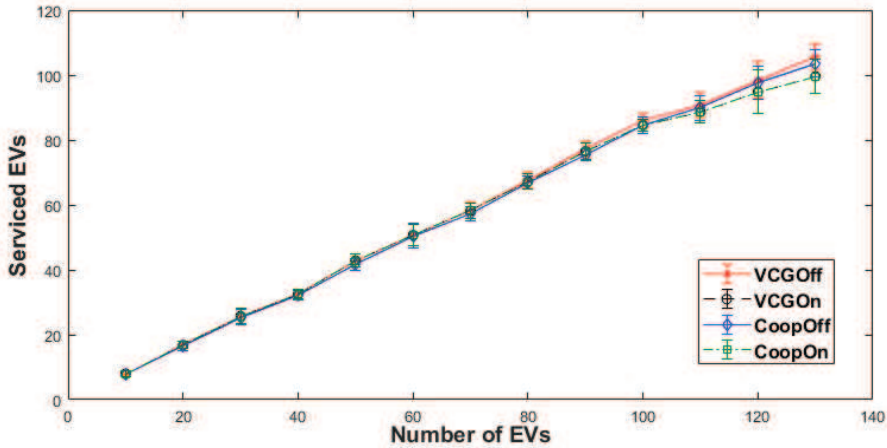

**Figure 3** Average number of serviced EVs

The utility of the serviced EVs (Figure 4) increases linearly with the number of EVs for all mechanisms. Regarding VCG, the utility for the online version is slightly higher compared to the offline one and especially for settings with more than 100 EVs. This can be explained by the fact that the offline charges more EVs, thus increasing the competition for the resources, and for this reason it calculates higher prices for the EVs (i.e., when the number of EVs to be charged increase, the addition of one EV is more likely to lead to others not being charged and for this reason the prices calculated by the VCG mechanism are higher. See also Section 7.3). Also, note that the rate of increase for the utility of the VCG mechanism, slows down for high numbers of EVs. This is related to the fact that when the demand for the resources increases a lot, this mechanism calculates high prices for the agents. Regarding the Coop, the utility for both variations is almost the same and for large numbers of EVs higher compared to VCG Offline. However, despite the fact that the Coop mechanism provides higher utility, it is vulnerable to manipulation in the case where EVs report their types non-truthfully. Whereas, VCG cannot be affected from such behaviour, as the EVs would have a loss any time they report anything other than their true types (see Section 7.4).

7.3 EXP3: Price for the EVs and budget for the system

Regarding the average price that each EV pays to the mechanism (Figure 5), the VCG offline calculates similar prices compared to the VCG online for small numbers of EVs, but later the prices increase rapidly. This can be explained by the fact that for each market clearing the competition for the resources is smaller as slightly less EVs are charged and for this reason the mechanism calculates lower prices. In terms of the Coop and given that it is a fixed price mechanism, the average price mainly depends on the number of serviced EVs.

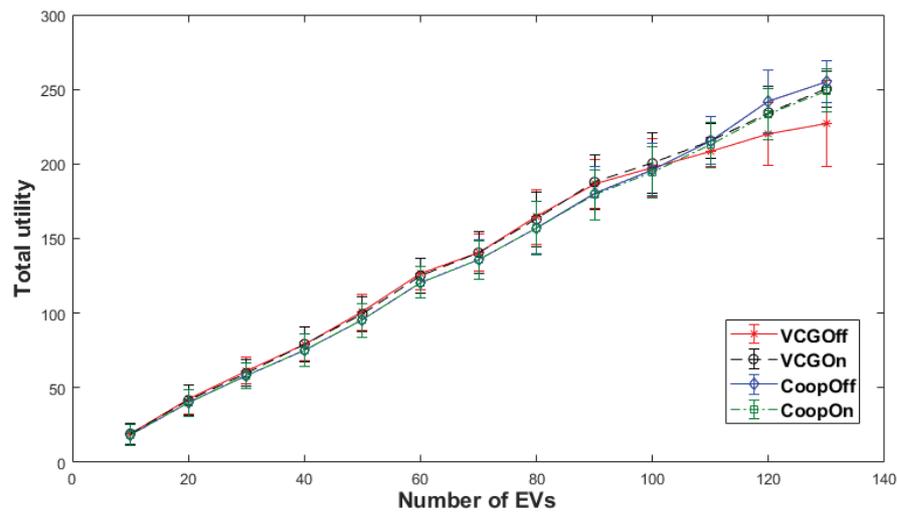

**Figure 4** Average total utility

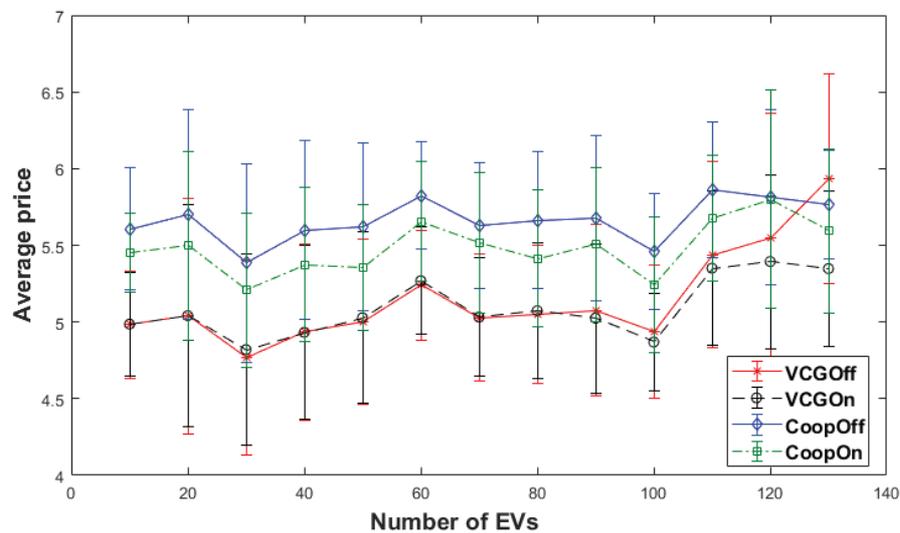

**Figure 5** Average price paid by the EVs

As long as the profit is concerned (Figure 6), for both mechanisms the offline variant leads to higher profit. Moreover, when comparing the VCG and the Coop mechanisms the profit is initially higher for the Coop, but when the competition for the resources increases a lot (i.e., more than 100 EVs) the VCG achieves higher profit. The online variants of the mechanisms follow a similar trend. As was expected, the results related to the profit of the mechanisms echo those related to the prices paid by the EVs. However, part of the higher profit of VCG is due to the fact that this mechanism achieves higher reduction of the imbalance cost. This is so, as in VCG all EVs that are selected to charge pay prices lower than their valuations and for this reason none of them leaves the mechanism without charging. However, in the case of Coop, some EVs leave as the price to pay is higher than their valuation. For this reason the Coop mechanism pays a higher imbalance cost especially for large numbers of EVs.

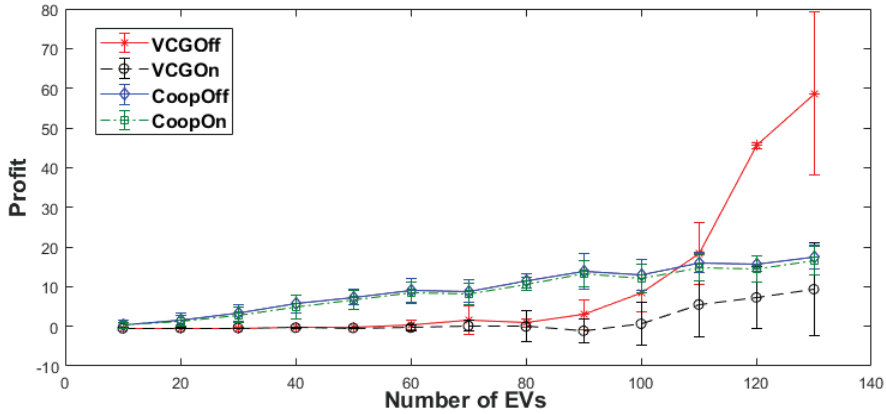

**Figure 6** Total profit for the mechanism

An interesting question is how the revenue changes with the number of charging stations. As can be seen in Figure 7, in a setting where the number of EVs is fixed to 80 and the charging stations vary from 6 to 18, the revenue of the VCG mechanism decreases with the number of stations. This can be explained by the fact that when the number of stations is small, the competition for the resources is high and for this reason the mechanism calculates high prices. However, as the number of the stations increases, the competition decreases and so do the prices. Interestingly, for more than 14 stations, the revenue becomes negative. This can be explained by the existence of the imbalance cost: Given an equal expected demand for each station, when up to 14 of them exist, the actual demand usually overcomes the expected one. Thus, the majority of the EVs have a positive transfer and for this reason the profit for the stations is positive. However, for more than 14 stations, the expected demand is higher than the actual one, and for this reason the transfers of a number of EVs are negative (i.e., they receive a payment from the mechanism) as these EVs reduce the imbalance cost and for this reason their existence has a positive impact in the system. This finding can be used to decide on the optimal number of charging stations for an area or a city. At the same time, the Coop calculates similar prices for all cases.

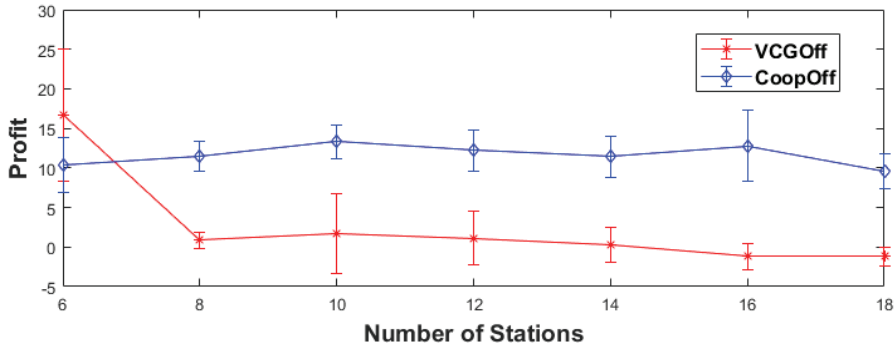

**Figure 7** Total profit - Variable number of stations

7.4 EXP4: Truthful VS non-truthful reporting

In the experiments presented so far, all EVs are assumed to report their types truthfully. However, this may not be always the case. Thus, here we evaluate a setting where some EVs report their valuations non-truthfully. By reporting lower valuation, many of these EVs are not selected to charge and get zero utility. Thus, it is obviously a negative choice for them and for this reason we focus in the case where some EVs report higher valuation. In particular, we assume that 10% of the EVs report 80% higher valuation.[6] In this case and as far as VCG is concerned, the non-truthful EVs have a 9.09% decrease in their utility despite the fact that 8.69% more of them are selected to charge. This happens due to the fact that the mechanism calculates higher prices for them. At the same time, the truthful EVs have a 1.35% decrease in their utility due to the fact that 0.46% less of them are charged. Thus, in the case of VCG, the agents do not have an incentive to misreport their valuation. These results confirm the theoretical evaluation presented in Section 5.2.

In contrast, when the Coop mechanism is used, the non-truthful EVs increase their utility by 6.07% as 10.67% more of them are selected to charge, while the truthful EVs face a 1.2% decrease in their utility due to the fact that 1.67% less of them are charged. Given this, EVs, which are rational agents, have an incentive to misreport their valuations. For this reason we further study the Coop mechanism in a scenario where all EVs lie about their valuations. In this case, the utility of all EVs is reduced by 1.57% due to the fact that $3,23\%$ less EVs are charged. Note that, due to the fact that all valuations shift upwards by 80%, the decrease in the total utility is not very high. Despite the fact that in Coop when all agents lie they receive a worse utility, and given the fact that if only some agents lie the rest receive a worse payoff, the agents will be locked to the lying strategy, as in the prisoners dilemma game. At the same time, the profit for the station is increased by 8.57% due to the fact that the higher valuations give the option to the mechanism to select fewer EVs but with high valuations and reduce the imbalance cost. However, given the fact that the municipality stations' goal is not to maximize profit but social welfare, there is an incentive to use the VCG mechanism instead of the Coop. Note that in all cases the statistical significance of the results has been verified using t-tests.

---

[6] For this value the liers have been observed to achieve the higher utility in the Coop mechanism.

## 8 Conclusions and future work

In this work, we presented market-based techniques to solve the problem of scheduling EVs and allocating them to charging stations. We considered two approaches, and for each one we considered both the offline and an online variant. In the first approach we used a fixed price mechanism, while in the second approach, we used the well known VCG mechanism and we proved that truthtelling is the dominant strategy. We evaluated our algorithms in a realistic setting and we observed that both have good scalability as they scale to hundreds of agents and tenths of charging stations. Moreover, we observed that the VCG mechanism leads to higher revenue for the stations and lower utility for the EVs in cases where the stations are highly congested. However, it is proven to not be vulnerable to agents' strategic behaviour. Finally, both approaches achieve approximately 83% serviced EVs.

As far as future work is concerned, we aim to apply online mechanism design techniques for the same problem, while we also want to add V2G and V2V energy transfer so as to use the EVs' batteries as storage devices and increase energy utilization and customer satisfaction [15]. Moreover, we aim to use queuing theory in order to add the ability for the EVs to wait in a queue in the charging stations. Finally, it is crucial to investigate machine learning techniques in order to achieve good prediction in future demand and minimise the imbalance cost.